\documentclass[useAMS,usenatbib]{mn2e}
\usepackage{natbib}
\usepackage{amsmath,amsfonts}
\usepackage{epsfig}
\usepackage{multirow}
\usepackage{times}

\def\reff@jnl#1{{\rm#1\/}}

\def\aj{\reff@jnl{AJ}}                  
\def\araa{\reff@jnl{ARA\&A}}            
\def\apj{\reff@jnl{ApJ}}                        
\def\apjl{\reff@jnl{ApJ}}               
\def\apjs{\reff@jnl{ApJS}}              
\def\ao{\reff@jnl{Appl.Optics}}         
\def\apss{\reff@jnl{Ap\&SS}}            
\def\aap{\reff@jnl{A\&A}}                       
\def\apjl{\reff@jnl{ApJ}}               
\def\aapr{\reff@jnl{A\&A~Rev.}}         
\def\aaps{\reff@jnl{A\&AS}}             
\def\azh{\reff@jnl{AZh}}                        
\def\baas{\reff@jnl{BAAS}}              
\def\jrasc{\reff@jnl{JRASC}}            
\def\memras{\reff@jnl{MmRAS}}           
\def\mnras{\reff@jnl{MNRAS}}            
\def\pra{\reff@jnl{Phys. Rev. A}}         
\def\prb{\reff@jnl{Phys. Rev. B}}         
\def\prc{\reff@jnl{Phys. Rev. C}}         
\def\prd{\reff@jnl{Phys. Rev. D}}         
\def\prl{\reff@jnl{Phys. Rev. Lett}}      
\def\pasp{\reff@jnl{PASP}}              
\def\pasj{\reff@jnl{PASJ}}              
\def\qjras{\reff@jnl{QJRAS}}            
\def\skytel{\reff@jnl{S\&T}}            
\def\solphys{\reff@jnl{Solar~Phys.}}    
\def\sovast{\reff@jnl{Soviet~Ast.}}     
\def\ssr{\reff@jnl{Space~Sci.Rev.}}     
\def\zap{\reff@jnl{ZAp}}                        
\def\nat{\reff@jnl{Nature}}             

\def\p#1by#2{{\partial{#1} \over \partial{#2}}}
\def\pp#1by#2#3{{\partial^2{#1} \over \partial{#2}\partial{#3}}}
\def\d#1by#2{{{\rm d}{#1} \over {\rm d}{#2}}}
\def\dd#1by#2#3{{{\rm d}^2{#1} \over {\rm d}{#2}{\rm d}{#3}}}

\title[Rocks in the Ring World]{The long wavelength view of GG\,Tau\,A: Rocks in the Ring World}

\label{firstpage}

\author[A.~M.~M.~Scaife]{
 Anna M. M. Scaife\thanks{email: a.scaife@soton.ac.uk}
 \vspace{0.03in}\\
 Department of Physics \& Astronomy, University of Southampton, Highfield, Southampton, SO17 1BJ\\
}

\date{Accepted ---; received ---; in original form \today}

\pagerange{\pageref{firstpage}--\pageref{lastpage}}

\pubyear{2010}

\begin{document}
\maketitle

\begin{abstract}
We present the first detection of GG\,Tau\,A at centimeter-wavelengths, made with the Arcminute Microkelvin Imager Large Array (AMI-LA) at a frequency of 16\,GHz ($\lambda=1.8$\,cm). The source is detected at $>6\,\sigma_{\rm rms}$ with an integrated flux density of $S_{\rm 16\,GHz} = 249\pm45\,\mu$Jy. We use these new centimetre-wave data, in conjunction with additional measurements compiled from the literature, to investigate the long wavelength tail of the dust emission from this unusual proto-planetary system. We use an MCMC based method to determine maximum likelihood parameters for a simple parametric spectral model and consider the opacity and mass of the dust contributing to the microwave emission. We derive a dust mass of $M_{\rm d}\approx0.1$\,M$_{\odot}$, constrain the dimensions of the emitting region and find that the opacity index at $\lambda>7$\,mm is less than unity, implying a contribution to the dust population from grains exceeding $\approx$4\,cm in size.
We suggest that this indicates coagulation within the GG\,Tau\,A system has proceeded to the
point where dust grains have grown to the size of small rocks with dimensions
of a few centimetres. Considering the relatively young age of the GG\,Tau association in combination with the low derived disk mass, we suggest that this system may provide a useful test case for rapid core accretion planet formation models.

\end{abstract}


\section{Introduction}
\label{sec:intro}

The evolution of protoplanetary disks is crucial for distinguishing between different models of planet formation, and the long wavelength emission from such disks is key to understanding this evolution. A number of theoretical models for planet formation within circumstellar disks have been proposed: the formation of gas giants by core accretion and atmospheric accumulation (Pollack et~al. 1996; Hubickyj et~al. 2005), is known to suffer from time scale difficulties, although these can be alleviated by the extended models such as that of Alibert et~al. (2005) who build on the standard core accretion models to provide a more rapid method of planet formation; alternatively the theory of planet formation by gravitational instability mechanisms (Boss et~al. 1992; 1997; Rice et~al. 2005) is proposed, although this method has been noted to require proto-planetary disk masses generally in excess of those currently measured from sub-mm data to proceed successfully. 

The second of these issues, that of missing mass, is exacerbated by an inexact understanding of the evolution of the dust emissivity, $\kappa_{\nu}$, in these objects. Disk masses derived from sub-mm and centimetre flux densities are critically sensitive to the value of this parameter, with $M_{\rm d}\propto \kappa_{\nu}^{-1}$. An increase in the emissivity results in a decrease in the derived mass, and vice-versa. Emissivity is a function of grain size and composition in the disk and, although it can be modeled theoretically, observationally determining its value and behaviour as a function of frequency is not trivial.

The effect of such multiplicity is a further issue which has been suggested to influence planet formation. It has been found that approximately a quarter of known extra-solar planets are located in binary or poly-stellar systems (e.g. Mugrauer \& Neuh{\"a}user 2009) and this bias is generally considered to be due to the increased dynamical complexity of such systems which results in their inner regions being swept clear of material, allowing gas to flow through the gap and influence the mass gain of gas giant planets (Ayliffe \& Bate 2010).

Determining the dust characteristics of the protoplanetary disks around young stars, whether in multiple systems or otherwise, is a complicated process. At sub-millimetre wavelengths it is complicated by opacity effects, as well as contamination by envelope, rather than disk, dominated emission at shorter wavelengths. At microwave frequencies the emission from disks is thought to be largely optically thin, avoiding the former as well as the latter of these issues. However, the emission is significantly fainter than at sub-mm frequencies, and furthermore can be contaminated by additional radio emission from ionized gas in the vicinity of these objects (e.g. Owen, Scaife \& Ercolano 2013). 

Radio frequency measurements of disks provide a useful tool as the emission is expected to be optically thin at such long wavelengths. In addition, in the Rayleigh-Jeans region (${\rm h}\nu << {\rm k}_{\rm B} T_{\rm d}$) the opacity of the thermal emission from dust grains can be well approximated by a power law, $\kappa_{\nu} \propto \nu^{\beta}$. This index $\beta$ can be related to the spectral index of flux density measurements, here defined as $S_{\nu} \propto \nu^{\alpha}$, as $\beta \simeq (1+\Delta)\times(\alpha - 2)$, where $\Delta$ is the ratio of optically thick to optically thin emission (Beckwith et~al. 1990). In the sub-mm region, 350\,$\mu$m to 1.3\,mm, this has been determined to have an average value $\Delta\simeq 0.2$ (Rodmann et~al. 2006; Lommen et~al. 2007), however at much longer cm-wavelengths $\Delta \rightarrow 0$, as the emission is entirely optically thin and consequently the largest grain sizes can be determined directly from a measure of the radio spectral index.

In spite of these advantages, there are also a number of disadvantages in observing disks at longer wavelengths. Firstly, the emission is fainter due to the spectrum falling off steeply at lower frequencies. Secondly, there is confusion at longer wavelengths from the additional radio emission often associated with protostellar systems. Although higher resolution imaging offers significant advantages in this regard, separating residual radio emission from the dust emission still requires a careful disentangling of spectral components (see e.g. Melis et~al. 2011).
It has been shown (AMI Consortium: Scaife et~al. 2012) that considering radio and dust components separately can give substantially different values for both the spectral slope of each component, important for determining opacity indices, as well as the normalization of each component, important for determining disk mass. Degeneracies between different parameters, increased in the case of lack of data or unfavourable spectral combinations may be a limiting factor in the accuracy with which such parameters can be determined.

In this paper we examine the long wavelength emission from the well-known GG\,Tau\,A disk/ring system, extending the SED to centimeter wavelengths; we briefly describe this system in \S~\ref{sec:ggtau}. In \S~\ref{sec:obs} we describe the new observations presented here and the data reduction method; in \S~\ref{sec:sed} we describe our SED fitting procedure and present the results of those fits; in \S~\ref{sec:disc} we discuss the implications of these results and in \S~\ref{sec:conc} we draw our conclusions.

\section{GG\,Tau\,A}
\label{sec:ggtau}

GG\,Tau is a quadruple system (Leinert et~al. 1991) within the Taurus molecular cloud (assumed distance 140\,pc). It consists of a pair of binary systems, denoted A and B. GG\,Tau\,A, the subject of this work, is classified as a Classical T Tauri Star (CTTS) and is itself a close binary system with a separation of 35\,AU. This binary pair is surrounded by a circumbinary disk (Dutrey et~al. 1994; Guilloteau et~al. 1999), which itself hosts a dense inner `ring' (Pi{\'e}tu et~al. 2011) of material with a width of only 50\,AU, which has lead to its nickname of ``the ring world" (Guilloteau et~al. 1999). The inner $\sim$150\,AU around the binary itself has been swept clear through the tidal interaction of the constituent stellar pair, although filamentary ``streamers" of material have been observed to extend from the inner edge of the outer circumbinary ring across this gap towards the central binary system and its inner circumstellar disk (McCabe et~al. 2002; Krist et~al. 2005; Pi{\'e}tu et~al. 2011).

In the centimetre-wave regime Bieging, Cohen \& Schwartz (1984) previously observed the GG\,Tau system at 2 \& 6\,cm, but with a 3\,$\sigma$ detection limit of 0.7\,mJy\,beam$^{-1}$ only GG\,Tau/N was detected in their data. Subsequently, the longest wavelength measurement made of GG\,Tau was at 7\,mm (Rodmann et~al. 2006) who detected extended emission at a resolution of 1.5\,arcsec, but did not investigate the source further. Simon \& Guilloteau (1993), observing the system at 2.6\,mm, noted that the mass of the system, assuming completely optically thin emission from a uniform dust disk would be 0.15\,M$_{\odot}$. For this estimate they assumed an isothermal disk, with a temperature of 15\,K. Using temperature estimates from $^{13}$CO data, resulting in $T({\rm 300\,AU})=20$\,K, and modeling the resolved sub-mm data, the dust mass in the disk+ring system was found to be 0.12\,M$_{\odot}$ (Guilloteau et~al. 1999) whilst the mass of the ring alone was found to be 0.09\,M$_{\odot}$ (Pi{\'e}tu et~al. 2011). Both estimates were somewhat smaller than that of Ohashi et~al. (1991) who derived a dust mass for the disk+ring system of 0.16\,M$_{\odot}$. This discrepancy may be explained by the surface density assumed by Ohashi et~al. (1991) which is approximately a factor of 1.6 higher than that used by the previously listed estimates. Allowing for differences in assumed physical characteristics, the range of dust mass estimates for GG\,Tau\,A is less than a factor of 2. Compared to the dynamical mass of the system (Guilloteau et~al. 1999) the dust is approximately $10\%$  as massive.

\section{Observations \& Imaging}
\label{sec:obs}

AMI comprises two synthesis arrays, one of ten 3.7\,m
antennas (SA) and one of eight 13\,m antennas (LA),
both sited at the Mullard Radio Astronomy Observatory at Lord's Bridge, Cambridge (AMI Consortium: Zwart
et~al. 2008). The telescope observes in 
the band 13.5--17.9\,GHz with eight 0.75\,GHz bandwidth channels. In practice, the two
lowest frequency channels (1 \& 2) are not generally used due to a lower response in this frequency range and interference from geostationary
satellites. The data in this paper were taken with the AMI Large Array (AMI-LA).

Observations of GG\,Tau were made with the AMI-LA during August 2011. AMI-LA data reduction is performed using the local software tool \textsc{reduce}. This applies
both automatic and manual flags for interference, 
shadowing and hardware errors, Fourier transforms the correlator data to synthesize frequency
channels and performs phase and amplitude
calibrations before output to disc in \emph{uv} FITS format suitable for imaging in
\textsc{aips}\footnote{\tt http://www.aips.nrao.edu/}. Flux (primary) calibration is performed using short observations of 3C286, 3C48 and 3C147. We assume I+Q flux densities for these sources in the
AMI-LA channels consistent with the updated VLA calibration scale (Perley \& Butler 2013). Since the AMI-LA measures
I+Q, these flux densities 
include corrections for the polarization of the calibrator sources. A correction is
also made for the changing air mass over the observation. From
other measurements, we find the flux calibration is accurate to better than
5 per cent (AMI Consortium: Scaife et~al. 2008; AMI Consortium:
Hurley-Walker et~al. 2009). Additional phase (secondary) calibration is done using interleaved observations of
calibrators 
selected from the Jodrell Bank VLA Survey (JVAS; Patnaik et~al. 1992). After calibration, the phase is generally stable to
$5^{\circ}$ for channels 4--7, and
$10^{\circ}$ for channels 3 and 8. The FWHM of the primary beam of the AMI-LA is $\approx 6$\,arcmin at 16\,GHz. Due to their superior phase stability only channels $4-7$ were used for this work, resulting in an effective total bandwidth of 3\,GHz and a central frequency of 15.7\,GHz, referred to in this work as 16\,GHz.

Reduced data were imaged using the AIPS data package. {\sc{clean}}
deconvolution was performed using the task 
{\sc{imagr}} which applies a differential primary beam correction to
the individual frequency channels to produce the combined frequency
image. Due to the low declination of this sample, uniform visibility weighting was used to improve the AMI-LA PSF. The AMI-LA is sensitive to angular scales from $\approx0.5-6$\,arcmin, although this varies as a function of declination, hour angle coverage and data flagging.

In what follows we use the convention: $S_{\nu}\propto \nu^{\alpha}$, where $S_{\nu}$ is
flux density (rather than flux, $F_{\nu}=\nu S_{\nu}$), $\nu$ is frequency and $\alpha$ is the spectral index. All errors quoted are 1\,$\sigma$.

\begin{figure*}
\centerline{\includegraphics[width=0.48\textwidth]{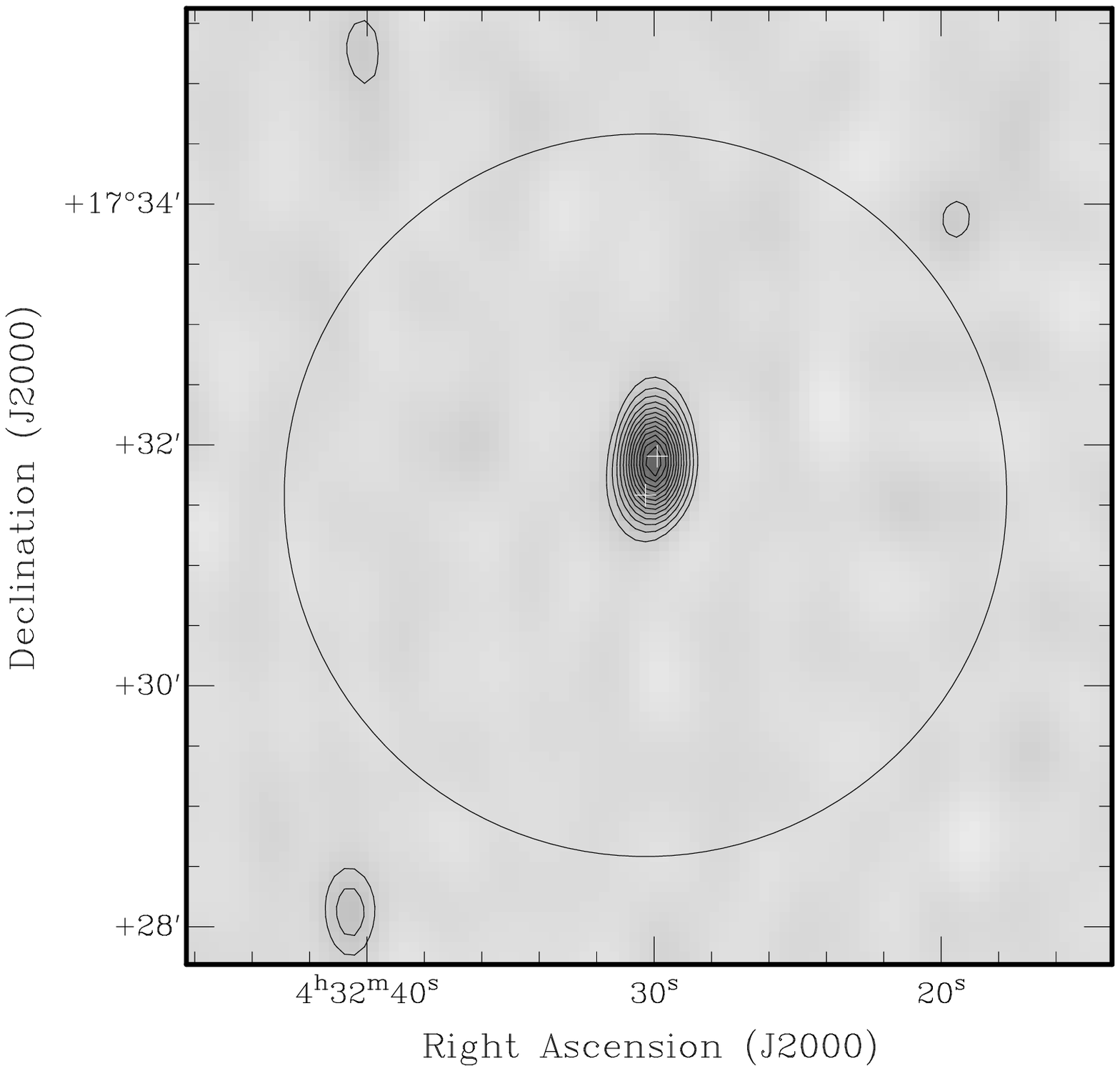}\qquad\includegraphics[width=0.48\textwidth, height=0.43\textwidth]{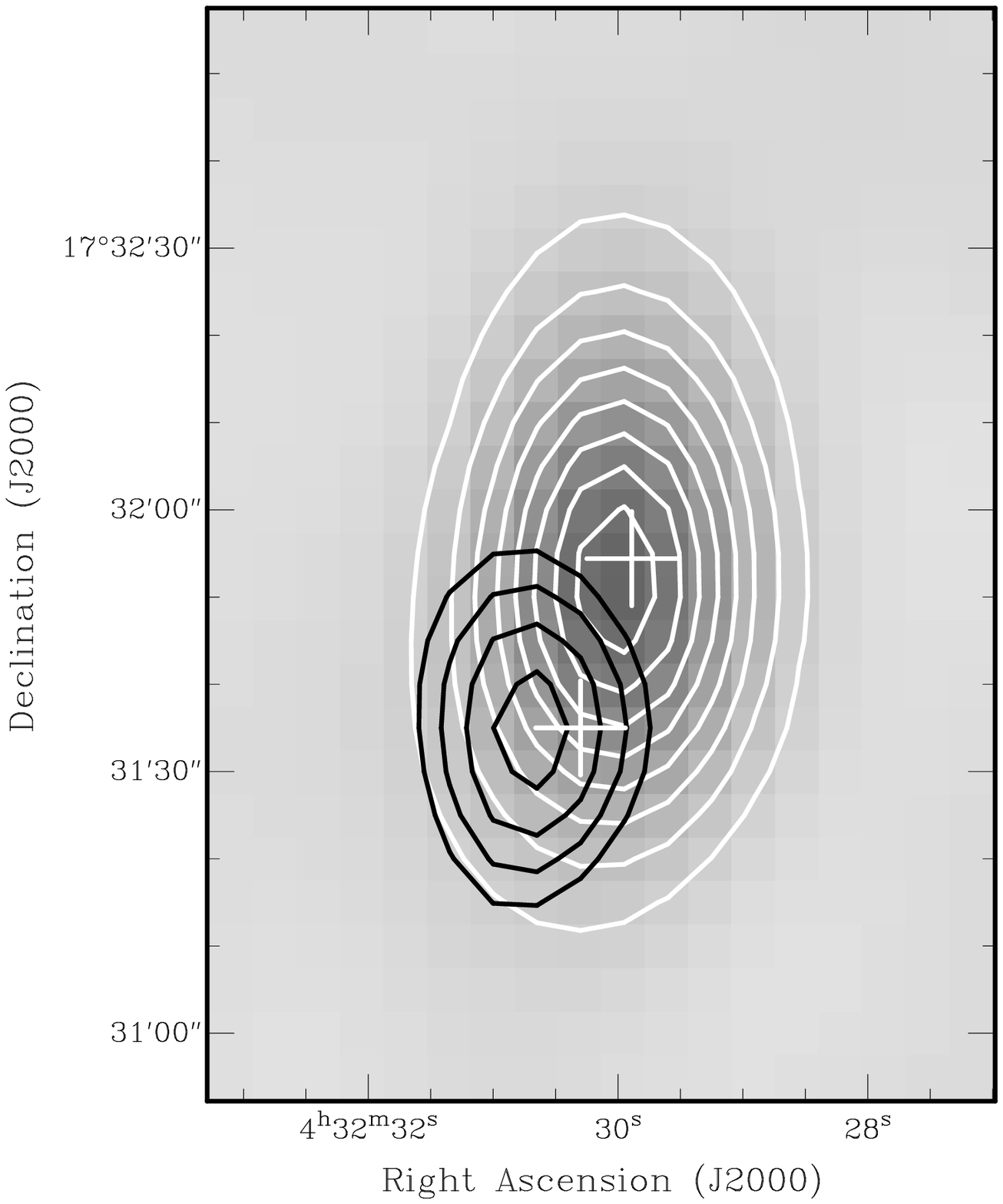}}
\centerline{(a)\hskip 0.48\textwidth (b)}
\caption{\textbf{GG\,Tau\,A}. (a) Greyscale and contours show original data from
the AMI-LA towards GG\,Tau\,A with linear contours from 5\,$\sigma$. The AMI-LA primary beam is indicated by a grey circle. (b) Greyscale and white contours show original data from
the AMI-LA towards GG\,Tau\,A at 5, 10, 20, 40\,$\sigma$ etc. Black contours show
uv-subtracted data, see text for details, at 3, 4, 5, 6\,$\sigma$. In both maps, white crosses
mark the positions of GG\,Tau/N to the North and GG\,Tau\,A to the South.\label{fig:img}}
\end{figure*}

\subsection{Separation of sources}

GG\,Tau\,A is closely separated from GG\,Tau/N, which is widely thought to be an extragalactic radio source. The AMI-LA data towards GG\,Tau is dominated by GG\,Tau/N, with the peak of the 1.8\,cm emission closely associated to the position of this radio source; an extension to the South can be seen in the direction of GG\,Tau\,A, see Fig.~\ref{fig:img}(a). We separate these sources by subtracting a point source located at the position of GG\,Tau/N directly from the visibilities. This subtraction assumes a flux density equivalent to the peak flux density in the AMI-LA map, and therefore a negligible contribution at this location from GG\,Tau\,A. The subtracted data reveals a point source at J\,04h32m30.7s~+17d31m35s, see Fig.~\ref{fig:img}(b). 

This point source is detected at $>6\,\sigma_{\rm rms}$ with a flux density of $S_{\rm 1.8\,cm} = 249\pm45\,\mu$Jy. The error on this flux density includes an increased uncertainty on the absolute calibration to a level of 10\,\% to account for additional offsets which may be introduced by the subtraction of GG\,Tau/N as well as a contribution from the thermal noise. The error is given as the quadrature sum of these quantities, $\sigma = \sqrt{(0.1\,S_{\rm 1.8\,cm})^2+\sigma_{\rm rms}^2}$. The synthesized beam of the AMI-LA towards GG\,Tau for these data is $\Omega = 44\times 25$\,arcsec, resulting in a positional accuracy for the AMI-LA centroid at an effective signal-to-noise of 5.5\,$\sigma$ of $\approx 7$\,arcsec. Consequently we identify this point source with GG\,Tau\,A as the radio peak is consistent with the infra-red position of GG\,Tau\,A to $\approx 1\sigma$. The infra-red position of GG\,Tau\,A is indicated by a cross in Fig.~\ref{fig:img}(b).

\subsection{Expected contamination by extragalactic radio sources}

At 16\,GHz we expect a certain number of extragalactic radio sources to be detected within the AMI-LA primary beam. We quantify this number using the 15\,GHz source counts model from de Zotti et~al. (2005) scaled to the 10C survey source counts (Davies et~al. 2010).  The rms noise in our dataset is $38\,\mu$Jy\,beam$^{-1}$ and from this model we predict that we should see $\simeq 1$ radio sources within a 6\,arcmin FWHM primary beam above a 5\,$\sigma$ flux density of 190\,$\mu$Jy. Our results are compatible with this prediction as GG\,Tau/N is classified as an extra-galactic source.

\section{Blackbody Spectrum}
\label{sec:sed}

The long wavelength SED of CTTSs is expected to be dominated by the thermal dust contribution from the circumstellar disk. This contribution is generally considered to be well described by a modified blackbody spectrum,
\begin{equation}
S_{\nu} \propto \nu^{\beta} B_{\nu}(T_{\rm d}),
\end{equation}
where $B_{\nu}(T_{\rm d})$ is the Planck function at a temperature $T_{\rm d}$, and $\beta$ is the opacity index such that the dust opacity $\kappa_{\nu}$ can be described as $\kappa_{\nu} = \kappa_0 (\nu/\nu_0)^{\beta}$. 

In the simplest interpretation, for thermal dust emission the flux density at a given frequency is related to the disk mass by
\begin{equation}
M_{\rm d} = \frac{S_{\nu}\Psi d^2}{\kappa_{\nu} B_{\nu}(T_{\rm d})},
\end{equation}
where $\Psi$ is the dust to gas ratio, $\kappa_{\nu}$ is the opacity at frequency $\nu$ and $B_{\nu}(T_{\rm d})$ is the value of the Planck function at frequency $\nu$ for a dust temperature $T_{\rm d}$. We use the dust opacity, $\kappa_{\nu} = 0.9(\nu/90.9\,{\rm GHz})^{\beta}$\,cm$^2$\,g$^{-1}$ (Lommen et~al. 2007).

Consequently the modified blackbody spectrum of a given object can be normalized at a frequency $\nu_0$ by a given disk mass, $M_{\rm d}$ such that
\begin{equation}
\label{eq:mbb}
S_{\nu} = \frac{M_{\rm d}}{\Psi d^2} \kappa_0 (\nu/\nu_0)^{\beta} B_{\nu}(T_{\rm d}).
\end{equation}

\begin{figure}
\centerline{\includegraphics[width=0.48\textwidth]{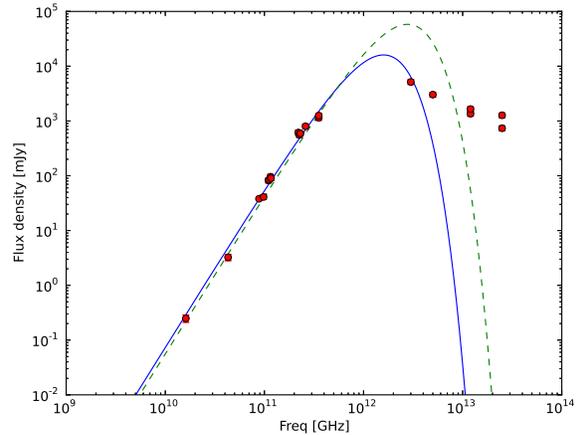}}
\caption{SED of GG\,Tau\,A. Flux density measurements from this work and compiled from the literature, see Table~\ref{tab:data}, are shown as data points with associated error bars. The maximum likelihood SED is shown as a solid line, see \S~\ref{sec:sed}, and an SED with the dust temperature fixed at $T_{\rm d}=35$\,K is shown as a dashed line, see \S~\ref{sec:sed} for details. \label{fig:sed}}
\end{figure}

In order to fit an SED to GG\,Tau\,A we compile a set of data at different frequencies from the literature. These data are listed in Table~\ref{tab:data} and shown in Fig.~\ref{fig:sed}. These data are selected to only include measurements in which the disk and ring are expected to be spatially unresolved, i.e. the disk has not been resolved out as is the case in (e.g) Pi{\'e}tu et~al. (2011). We also restrict this compilation to that region of the spectrum expected to be dominated by the thermal emission from the disk and ring, limiting our coverage to $\lambda>10\,\mu$m. Where no error is stated on these archival measurements we adopt a standard 10\,\% error on the flux density. We fit the function in Eq.~\ref{eq:mbb} to these data with dust temperature, disk mass and opacity index as free parameters and using a normalizing frequency of $\nu_0 = 100$\,GHz. We use a metropolis hastings MCMC code to determine the maximum likelihood parameter values, which we find to be $T_{\rm d, ML} = 19.42\pm0.55$\,K, $M_{\rm d, ML} = 0.099\pm0.004$\,M$_{\odot}$ and $\beta_{\rm ML} = 0.96\pm0.04$. The fitted model is shown in Fig.~\ref{fig:sed}. As expected there is a strong negative correlation between the opacity index and dust temperature, with a correlation co-efficient of $p=-0.82$, as well as weaker correlations between disk mass and dust temperature ($p=-0.64$) and disk mass and opacity index ($p=0.33$).

\begin{table}
\caption{Flux densities for GG\,Tau\,A from this work and compiled from the literature. Column [1] lists the frequency of the measurement; column [2] lists the flux density in mJy; column [3] lists the reference for each measurement. \label{tab:data}}
\begin{center}
\begin{tabular}{lll}
\hline\hline
$\nu$ & $S_{\nu}$ & ref. \\

 [GHz] & [mJy] & \\
\hline
16.1   &   $0.249\pm0.045$ & this work  \\
43.0 &     $3.24 \pm	0.42$ & Rodmann et~al. 2006\\
88.2   &     $38\pm2$        & Guilloteau et~al. 1999\\
98.0  &     $41 \pm      4.1$  & Ohashi et~al. 1991\\      
110.0  &     $82\pm        8.2$  & Koerner et~al. 1993\\
111.1   &      $85\pm5$ 	& Guilloteau et~al. 1999\\
115.3   &    $96 \pm      9.6$  & Kawabe et~al. 1993\\ 
116.3 &      $90\pm       9.0$  & Simon et~al. 1992\\ 
214.3   &     $604\pm8$		& Guilloteau et~al. 1999\\
220.0 &    $615 \pm       61.5$ & Koerner et~al. 1993\\ 
225.0  &    $557\pm3$           & Harris et~al. 2012 \\
230.0 &    $ 593\pm      53$ & Beckwith et~al. 1990\\ 
230.0 &    $593\pm59$    & Andrews \& Williams 2005 \\ 
260.0 &    $ 800\pm       80$ & Beckwith et~al. 1991\\ 
353.0 &    $1250\pm          125$ & Johnstone et~al. 1993\\ 
353.0	&    $1255\pm	125.5$	& McCabe et~al. 2006\\	
3000.0	&    $5160\pm	520$ & IPAC 1986\\
5000.0	&    $3030\pm	303$& IPAC 1986\\
12000.0	&    $1360\pm	140$ & Cutri et~al. 2012\\
12000.0	&    $1650\pm	165$& IPAC 1986\\
25000.0	&    $736\pm	80$ & Cutri et~al. 2012\\
25000.0	&    $1270\pm	130$& IPAC 1986\\
\hline
\end{tabular}
\end{center}
\end{table}

\subsection{Disk Mass and Dust Temperature} 
\label{sec:diskmass}

The disk mass recovered by fitting the GG\,Tau\,A SED is intermediate to the values recovered previously for the dense inner ring, $M_{\rm ring}=0.09$\,M$_{\odot}$ (Pi{\' e}tu et~al. 2011), and the larger circumbinary disk-ring system, $M_{\rm system}=0.12$\,M$_{\odot}$ (Guilloteau et~al. 1999). The dust temperature recovered from fitting the SED is relatively cold. Guilloteau et~al. (1999) suggested that the temperature of the GG\,Tau\,A system dropped steeply with radial distance from the central object, resulting in a temperature of $T(R=300\,{\rm AU})\approx 20$\,K. Since the single temperature assumed in Eq.~\ref{eq:mbb} will be a mass-weighted average temperature for the combined emission this suggests that the mass of dust contributing to the centimetre-wave flux density is dominated by the colder dust population.

At smaller radial distances the dust temperature of the GG\,Tau\,A system has been modeled to be much higher, with $T(R=50\,{\rm AU})\approx 35$\,K. An SED with a dust temperature fixed to this value is also fitted the dataset in Table~\ref{tab:data}. The best fit SED from this fitting is shown as a dashed line in Fig.~\ref{fig:sed} where it can be seen that it is inconsistent with the measured FIR shape of the GG\,Tau\,A SED.

\subsection{Opacity Index}
\label{sec:opacity}

Given the power-law approximation to the dust opacity as stated in \S~\ref{sec:sed}, the opacity index, $\beta$, can be determined using the ratio of flux densities at two frequencies such that
\begin{equation}
\frac{S_{\nu_1}}{S_{\nu_2}} = \gamma \left( \frac{\nu_1}{\nu_2} \right)^{3+\beta} \hskip .01\textwidth {\rm where} \hskip .01\textwidth \gamma = \frac{ \rm{e}^{h\nu_2/kT_{\rm d}} - 1}{ \rm{e}^{h\nu_1/kT_{\rm d}} - 1}
\end{equation}
and consequently
\begin{equation}
\label{eq:beta}
\beta = \frac{\ln\left(S_{\nu_1}/S_{\nu_2}\right) - \ln\gamma}{\ln\left({\nu_1}/{\nu_2}\right)} - 3
\end{equation}
with an associated error
\begin{equation}
\sigma_{\beta} =  \left|\frac{1}{ \ln\left( {\nu_1}/{\nu_2} \right) }\right|  \sqrt{   \left( \frac{\sigma_{S_{\nu_1}}}{S_{\nu_1}} \right)^2 + \left( \frac{\sigma_{S_{\nu_2}}}{S_{\nu_2}} \right)^2  }.
\label{eq:sigb}
\end{equation}

Fitting all of the data in Table~\ref{tab:data} with a single modified blackbody results in an opacity index $\beta=0.96\pm0.04$, see \S~\ref{sec:sed}. However the composite nature of the GG\,Tau\,A system, with its disk plus ring, suggests that although the data seem well modeled by a single modified blackbody there may be multiple components contributing to the SED. Specifically we would expect the longest wavelength emission to have a larger relative contribution from that dust population which has the largest grain sizes and hence a flattened spectrum. Consequently, fixing the dust temperature at 20\,K we re-fit only those data with $\nu<100$\,GHz. This fit results in very similar parameter constraints: $M_{\rm disk} = 0.096\pm0.005$\,M$_{\odot}$ and $\beta=1.07\pm0.10$. However, the allowed range of parameter space, compared with the full SED fit, allows a range of much smaller disk masses and flatter values of $\beta$. Furthermore, using the expressions in Eqs.~\ref{eq:beta}~\&~\ref{eq:sigb}, the opacity index from 1.8\,cm to 7\,mm is $\beta_{\rm 1.8\,cm}^{\rm 7\,mm} = 0.64\pm0.24$, which implies a much flatter spectrum. This suggests that if there is a separate contribution from a larger dust grain population it only appears at wavelengths longer than $\sim$7\,mm. These results are summarized in Table~\ref{tab:res}.

\begin{table}
\begin{center}
\caption{Maximum likelihood parameter constraints from model fitting, see \S~\ref{sec:sed} for details. \label{tab:res}}
\begin{tabular}{lccc}
\hline\hline
Data range & $M_{\rm d}$ & $\beta$ & $T_{\rm d}$ \\
           & [M$_{\odot}$] &       &   [K] \\
\hline
\emph{Prior} & [0, 1] & [0, 2] & [0, 50]\\
\hline
All data & $0.099\pm 0.004$ & $0.96\pm0.04$ & $19.42\pm0.55$ \\
$\lambda > 3$\,mm & $0.096\pm0.005$ & $1.07\pm0.10$ & $20^{\dagger}$ \\
$\lambda \ge 7$mm & $-$ & $0.69\pm0.26$ & $-$ \\
\hline
\end{tabular}
\begin{minipage}{0.5\textwidth}{
$^{\dagger}$ fixed parameter.
}
\end{minipage}
\end{center}
\end{table}

\section{Discussion}
\label{sec:disc}

\subsection{Optically thick fraction}

In the Rayleigh-Jeans region the thermal emission from dust grains can be well approximated by a power law with $S_{\nu} \propto \nu^{\alpha}$. The spectral index $\alpha$ is related to the dust opacity index $\beta$ as
\begin{equation}
\label{eq:beta2}
\beta \simeq (1+\Delta)\times(\alpha - 2),
\end{equation}
where $\Delta$ is the ratio of optically thick to optically thin emission (Beckwith et~al. 1990). In the sub-mm region, 350\,$\mu$m to 1.3\,mm, this has been determined to have an average value $\Delta\simeq 0.2$ (Rodmann et~al. 2006; Lommen et~al. 2007). 

Although at longer wavelengths $\Delta$ is expected to tend to zero, the exact value is a function of the column density, which may be very high in the inner regions of the disk causing the opacity to be non-negligible even at centimetre wavelengths. It may be calculated using the radial power-law indices of the disk surface density and temperature, denoted $p$ and $q$, respectively, as 
\begin{equation}
\Delta \approx -p \times \left[(2-q) \ln\{(1-p/2)\bar{\tau}\}\right]^{-1}
\end{equation}
where $\bar{\tau}$ is the average disk opacity at the measurement frequency, defined as
\begin{equation}
\label{eq:tau}
\bar{\tau} \equiv \frac{\kappa_{\nu} M_{\rm d}}{\Psi \cos \theta \pi R_{\rm d}^2}
\end{equation}
(e.g. Beckwith et~al. 1990), where $\theta$ is the inclination of the disk to the line of sight ($\theta=0^{\circ}$ for a face on disk) and $R_{\rm d}$ is the outer radius of the disk.

The surface density of disks is expected to decrease with radius, leading to positive values of $p$. A value of $p=1.5$ is typically assumed (Beckwith et~al. 1990; Rodmann et~al. 2006) and we also adopt this value here. The power-law index of the radial temperature, defined such that $T(r) = T_0(r/r_0)^{-q}$, is uniquely determined from the infra-red spectral index, $\alpha_{\rm IR}$, calculated at wavelengths where the opacity is expected to be high, typically $\lambda < 100\,\mu$m, such that $q = 2/(3-\alpha_{\rm IR})$. Although we note that this power-law approximation breaks down at shorter wavelengths where reprocessing of stellar photons will modify the value of $q$. Using measurements from the IRAS and WISE catalogues, see Table~\ref{tab:data}, we find that $q = 0.5-0.6$ for GG\,Tau\,A. However, we note that the more complex modeling of this system by Guilloteau et~al. (1999) found a value of $q=0.9\pm0.1$, implying a steeper radial temperature power-law.

\begin{figure}
\centerline{\includegraphics[width=0.48\textwidth]{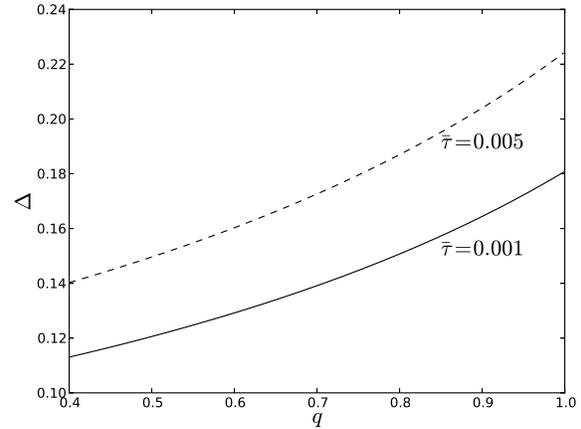}}
\caption{Fraction of optically thick to optically thin emission, $\Delta$, at $\lambda=2$\,cm as a function of the radial temperature power-law index, $q$, for limiting values of the average disk opacity, $\bar{\tau}$, see \S~\ref{sec:opacity} for details.  \label{fig:q_v_delta}}
\end{figure}

At millimetre-wavelengths, Lommen et~al. (2007) adopted a value of $\bar{\tau}=0.02$ at 3.3\,mm and Rodmann et~al. (2006) adopted $\bar{\tau}=0.01$ at 7\,mm, consistent with the results of Beckwith et~al. (1990). Consequently, at 1.8\,cm we expect that $\bar{\tau}\ll 0.01$. Extrapolating linearly from the millimetre values, gives a value of $\bar{\tau}=3.7\times 10^{-3}$ at a wavelength of 1.8\,cm. Since this is an inexactly determined value we consider a range of $0.001\le \bar{\tau}\le 0.005$. Combining this with the adopted value of $p$ we find that $\Delta=0.12-0.15$ for $q=0.55$ and $\Delta=0.16-0.20$ for $q=0.9$. The general behaviour of $\Delta$ as a function of $q$ and $\bar{\tau}$ is illustrated in Fig.~\ref{fig:q_v_delta}. 

Using data at 1.8\,cm and 7\,mm we find a spectral index of $\alpha_{0.7}^{1.8}=2.61\pm0.23$. The error on this index is a function not only of the precision with which the flux densities are measured at each frequency but also the bandwidth between those frequencies, with more closely separated frequencies necessarily resulting in larger errors on spectral index. We note that for the available data at 7\,mm and 1.8\,cm the error on spectral index is an order of magnitude larger than the spread in the values of $\beta$ produced by the uncertainty in $\Delta$ allowed by our range of $\bar{\tau}$ values. Using Equation~\ref{eq:beta2}, we find $\beta = 0.69\pm0.26$. The indicative error on this value comes from propagating both the error on the spectral index as well as an error from treating the spread in allowed values of $\Delta$ as one sigma bounds on a central value. This value for $\beta$ is consistent with that found from Eqs.~\ref{eq:beta}, see \S~\ref{sec:opacity}.

\subsection{Grain size distribution}

The opacity index at a wavelength, $\lambda$, is expected to be related to the opacity index of the diffuse interstellar medium (ISM) as $\beta(\lambda) \approx (\chi-3)\beta_{\rm ISM}$, where $\chi$ is the power-law index of the grain size distribution such that ${\rm d}n/{\rm d}a \propto a^{-\chi}$. Grain sizes $a$ are limited in size to a maximum of $a \le a_{\rm max}$. For the typically assumed MRN power-law distribution (Mathis, Rumpl \& Nordsieck 1977), where $\chi=3.5$, an opacity index measured at a given wavelength, $\lambda$, with a value of $\beta(\lambda) \approx 1$ indicates $a_{\rm max} > \lambda$ and $\beta(\lambda) \le 1$ corresponds to $a_{\rm max} > 3\lambda$ (Draine 2006). Although it has been shown through simulations that the growth of dust particles in protoplanetary disks may result in a flatter power-law size distribution with $\chi=3$ (Tanaka et~al. 2005), Draine (2006) argued that a power-law index of $\chi=3.5$ was likely for protoplanetary environments where processes of fragmentation and coagulation were in competition to provide the grain size distribution, and that the maximum grain size would be determined by the extent to which coagulation had progressed.  

From these new data presented here we find that $\beta$ is consistent with unity into the centimeter regime implying that for $\chi=3.5$, $a_{\rm max} > 2$\,cm. However, the opacity index constrained from data at $\lambda>7$\,mm alone is less than unity. Taking the wavelength of measurement as the geometric mean of 7\,mm and 1.8\,cm suggests that $a_{\rm max} > 3\lambda \approx 4.2$\,cm. Once again this is consistent with a grain size distribution power-law index of $\chi=3.5$; assuming  $\beta_{\rm ISM}=1.7$ (Finkbeiner et~al. 1999), the measured opacity index implies a power-law grain size distribution with $\chi=3.4\pm0.2$. This implies that coagulation within the GG\,Tau\,A system has proceeded to the point where dust grains have grown to size of small rocks with dimensions of a few centimetres.

\subsection{Emitting Region}

\begin{figure}
\centerline{\includegraphics[width=0.48\textwidth]{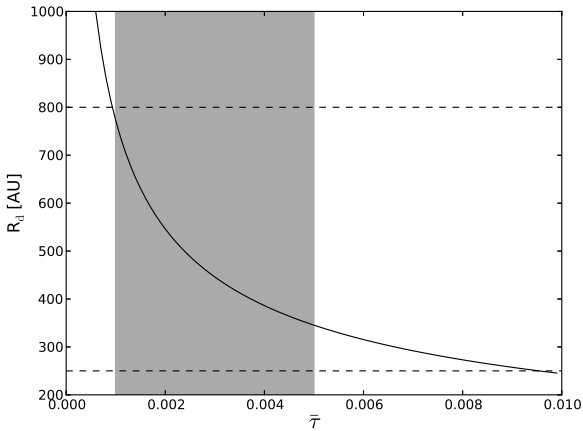}}
\caption{Outer radius of the emitting disk, $R_{\rm d}$, at 1.8\,cm as a function of the average disk opacity, $\bar{\tau}$. The proposed range of $\bar{\tau}$ at 1.8\,cm is indicated by the shaded area. Dashed lines indicate the proposed outer radius of the GG\,Tau\,A ring at 250\,AU and the outer radius of the GG\,Tau\,A disk at 800\,AU (Guilloteau et~al. 1999).  \label{fig:radius}}
\end{figure}

By considering Equation~\ref{eq:tau} we can use the disk mass calculated from the GG\,Tau\,A SED to calculate the emitting region by constraining its outer radius, $R_{\rm d}$. For a disk mass of $M_{\rm d}=0.099\pm0.01$\,M$_{\odot}$, see \S~\ref{sec:sed}, and assuming $\theta=37$\,degrees (Guilloteau et~al. 1999; Pi{\' e}tu et~al. 2011), a limiting $\bar{\tau}<0.01$ implies an outer radius of $R_{\rm d}>240$\,AU. For the proposed range of $\bar{\tau}= 0.001-0.005$ at $\lambda=2$\,cm the corresponding range of $R_{\rm d}$ is shown in Fig.~\ref{fig:radius}. 

These values encompass the radial range allowed by the model of Guilloteau et~al. (1999) who proposed that the GG\,Tau\,A system could be described by a dense ring of material with an outer radius of $R_{\rm ring} = 250\pm2$\,AU and a lower density disk with an outer radius of $R_{\rm disk}\approx 800$\,AU. The range of radii allowed by the mass and opacity constraints from the 1.8\,cm data suggest that the emission we are seeing is consistent with a combined contribution from both the ring and the disk. This is consistent with the mass calculated from the 1.8\,cm data being intermediate to that of the ring alone and the combined system. The value of the ring mass was previously calculated to be $M_{\rm ring}=0.09$\,M$_{\odot}$ (Pi{\' e}tu et~al. 2011), whereas the mass of the ring+disk system was calculated as $M_{\rm system}=0.12$\,M$_{\odot}$ (Guilloteau et~al. 1999).

\subsection{Possible contamination from other radio emission mechanisms}

At centimeter wavelengths contamination from thermal Bremsstrahlung or other radio emission mechanisms is a possibility for systems such as GG\,Tau\,A. Free-free emission from young stellar objects may arise as a consequence of shock ionization in molecular outflows (e.g. Curiel et~al. 1989), although this is unlikely in the case of GG\,Tau\,A due to the absence of any detected molecular emission from a jet or outflow. Molecular hydrogen has been detected in the regions interior to the GG\,Tau\,A ring (Beck et~al. 2012), however this ro-vibrational emission is thought to be associated with an accretion streamer rather than shock excitation. Such emission may also be a result of stimulation by X-ray or UV flux from a central object, but in this case the distribution of the H$_2$ emission imaged at high resolution was found to be inconsistent with such a scenario (Beck et~al. 2012). Such X-ray or UV flux can also give rise to free-free emission from photo-ionization of disk winds in CTTSs (Owen, Scaife \& Ercolano 2013), however in the case of GG\,Tau\,A no X-ray counterpart is detectable ($L_{\rm X} < 2.5\times 10^{29}$\,erg\,s$^{-1}$;  Neuh{\"a}user et~al. 1995) and photo-ionization due to UV flux is also unlikely as this is expected to result in an isothermal flow with temperatures of approximately $10^4$\,K. For GG\,Tau\,A the $2-1\,S(1)/1-0\,S(1)$ line ratio indicates much lower temperatures of $\sim 1700\pm100$\,K (Beck et~al. 2012), inconsistent with this scenario.

Weak-line T~Tauri Stars (WTTSs) are expected to produce radio emission of a non-thermal nature due to magnetic activity in the stellar corona. This radio luminosity of such emission is tightly coupled to X-ray emission as a common energy reservoir is thought to fuel both the magnetic activity and heating of the surrounding gas giving rise to thermal Bremsstrahlung in the X-ray regime (see Feigelson \& Montmerle 1999 for a review). Such radio emission is expected to follow the `G{\"u}del-Benz relation' (G{\"u}del \& Benz 1993; G{\"u}del 2002), often represented as the ratio $L_{\rm X}/L_{\rm rad}\sim 10^{15}$. Given the limits on X-ray luminosity from the ROSAT satellite (Neuh{\"a}user et~al. 1995) this relation would suggest a radio flux density of $S<5\,\mu$Jy at $5-8$\,GHz for GG\,Tau\,A. Even assuming a flat spectral index to 16\,GHz this provides a negligible contribution to the measured flux density. CTTS systems are expected to be magnetically similar to WTTS, although non-thermal gyro-synchrotron emission produced close to the star is expected to suffer substantial absorption by disk and YSO winds and therefore appear reduced. If this is indeed the case the already negligible expected contribution from non-thermal emission to the 16\,GHz flux density will be reduced even further.

\subsection{Comparison to similar systems}
\label{sec:comp}
Grain growth such as that seen here in the GG\,Tau\,A system is not unknown elsewhere. Perhaps the proto-typical system for such spectral flattening at long wavelengths is TW\,Hya (Wilner et~al. 2000), where 7\,mm observations at high resolution reveal a disk, thought to be seen almost face on, with a $\beta \approx 1$. However at longer wavelengths, $\lambda=3.5$\,cm, a substantial excess of emission (confirmed as dust emission rather than free-free emission due to a non-detection at 6\,cm) is detected, which can only be explained by the presence of centimeter-size grains (Wilner et~al. 2005). 

The disk mass in the TW\,Hya system is similar to that derived here, $M_{\rm d}\approx 0.1$\,M$_{\odot}$, neither high enough for the gravitational instability method of planet formation to occur (Boss 1998). Noting this, we consider the respective ages of the two systems. The TW\,Hya system is substantially older (5-10\,Myr; Song et~al. 2003) than GG\,Tau\,A ($\sim 1.5$\,Myr; White et~al. 1999). This is important when considering the implications of grain growth and the potential of the system to form planetesimals via the core accretion model. The canonical timescale for core accretion planet formation is 8\,Myr (Pollack et~al. 1996), nominally precluding planet formation in both the GG\,Tau\,A and TW\,Hya systems. However, Hubickyi et~al. (2004) calculated that a reduction in dust opacity could half this timescale and, subsequently, the migration and disk evolution model of Alibert et~al. (2005) demonstrated that in fact the formation timescale could be reduced to as little as $1-3$\,Myr. Furthermore, this reduction in timescale due to migration is not mutually exclusive with that due to dust opacity reduction and the combination of effects would shorten formation times even further. These revised versions of the core accretion mechanism of planet formation suggest that planet formation is feasible even in systems as young as GG\,Tau\,A. Indeed, planet formation has been suggested for even younger systems still, such as in the case of HL\,Tau, a Class~I object ($10^5$\,yr; Greaves et~al. 2008). Although in the case of HL\,Tau the disk mass was found to be high enough to allow the gravitational instability model of planet formation to operate. Once again the opacity index of the long wavelength emission from HL\,Tau was suggestive of centimeter-size grains and in this case a surface density enhancement (4.5\,$\sigma$) was detected and proposed as a potential protoplanet.

\section{Conclusions}
\label{sec:conc}

We have detected the GG\,Tau\,A ring/disk system at a wavelength of $\lambda=1.8$\,cm. Using these new data in combination with archival data compiled from the literature we have investigated the long wavelength SED of the system with a number of results.

We propose that the detected flux is due to thermal dust emission from the GG\,Tau\,A system and that additional radio emission mechanisms contribute negligibly to the integrated flux density, if at all. We suggest that, although the centimetre-wave emission is dominated by the dense GG\,Tau\,A ring, the outer radius of the emission is larger than that of the ring alone and is therefore likely to encompass the complete ring/disk system. We find a dust mass for the GG\,Tau\,A combined ring/disk system of $M_{\rm d}=0.099\pm0.004$\,M$_{\odot}$.

We find that the opacity index of the GG\,Tau\,A long wavelength emission is broadly consistent with unity, but there are indications that the spectrum flattens further at wavelengths longer than 7\,mm. We suggest that this flattening of the spectrum implies a dust population with maximum grain sizes in excess of 4\,cm and that this indicates coagulation of dust grains within the GG\,Tau\,A system into bodies more similar in size to small rocks than traditional dust.

With the indication of grain growth in the disk, and considering the relatively young age of the GG\,Tau association ($\sim1.5$\,Myr) in combination with the low derived disk mass, we suggest that this may be a good system with which to test rapid planet formation models such as that of Alibert et~al. (2005).

\section{ACKNOWLEDGEMENTS}
We thank the anonymous referee whose useful comments improved this paper. We thank the staff of the Lord's Bridge Observatory for their
invaluable assistance in the commissioning and operation of the
Arcminute Microkelvin Imager. The AMI-LA is supported by Cambridge
University and the STFC.


\begin{thebibliography}{}
\setlength{\labelwidth}{0pt}

\bibitem[\protect\citeauthoryear{Alibert et 
al.}{2005}]{2005A&A...434..343A} Alibert Y., Mordasini C., Benz W., Winisdoerffer C., 2005, A\&A, 434, 343 

\bibitem[\protect\citeauthoryear{Andrews 
\& Williams}{2005}]{2005ApJ...631.1134A} Andrews S.~M., Williams J.~P., 2005, ApJ, 631, 1134 

\bibitem[\protect\citeauthoryear{Ayliffe 
\& Bate}{2010}]{2010MNRAS.408..876A} Ayliffe B.~A., Bate M.~R., 2010, MNRAS, 408, 876 

\bibitem[\protect\citeauthoryear{Beckwith et 
al.}{1990}]{1990AJ.....99..924B} Beckwith S.~V.~W., Sargent A.~I., Chini 
R.~S., Guesten R., 1990, AJ, 99, 924 

\bibitem[\protect\citeauthoryear{Bieging, Cohen, 
\& Schwartz}{1984}]{1984ApJ...282..699B} Bieging J.~H., Cohen M., Schwartz P.~R., 1984, ApJ, 282, 699 

\bibitem[\protect\citeauthoryear{Boss}{1992}]{1992ASPC...32..195B} Boss 
A.~P., 1992, ASPC, 32, 195 

\bibitem[\protect\citeauthoryear{Boss}{1996}]{1996ApJ...469..906B} Boss 
A.~P., 1996, ApJ, 469, 906 

\bibitem[\protect\citeauthoryear{Boss}{1998}]{1998ApJ...503..923B} Boss 
A.~P., 1998, ApJ, 503, 923 

\bibitem[\protect\citeauthoryear{Cutri 
\& et al.}{2012}]{2012yCat.2311....0C} Cutri R.~M., et al., 2012, yCat, 2311, 0 

\bibitem[\protect\citeauthoryear{Davies et 
al.}{2011}]{2011MNRAS.415.2708A} Davies M.~L., et al., 2011, MNRAS, 415, 
2708 

\bibitem[\protect\citeauthoryear{de Zotti et 
al.}{2005}]{2005A&A...431..893D} de Zotti G., Ricci R., Mesa D., Silva L., Mazzotta P., Toffolatti L., Gonz{\'a}lez-Nuevo J., 2005, A\&A, 431, 893 

\bibitem[\protect\citeauthoryear{Di Francesco et 
al.}{2008}]{2008ApJS..175..277D} Di Francesco J., Johnstone D., Kirk H., 
MacKenzie T., Ledwosinska E., 2008, ApJS, 175, 277 

\bibitem[\protect\citeauthoryear{Draine}{2006}]{2006ApJ...636.1114D} Draine 
B.~T., 2006, ApJ, 636, 1114 

\bibitem[\protect\citeauthoryear{Dutrey, Guilloteau, 
\& Simon}{1994}]{1994A&A...286..149D} Dutrey A., Guilloteau S., Simon M., 1994, A\&A, 286, 149 

\bibitem[\protect\citeauthoryear{Finkbeiner, Davis, 
\& Schlegel}{1999}]{1999ApJ...524..867F} Finkbeiner D.~P., Davis M., Schlegel D.~J., 1999, ApJ, 524, 867 

\bibitem[\protect\citeauthoryear{Greaves et 
al.}{2008}]{2008MNRAS.391L..74G} Greaves J.~S., Richards A.~M.~S., Rice 
W.~K.~M., Muxlow T.~W.~B., 2008, MNRAS, 391, L74 

\bibitem[\protect\citeauthoryear{Guilloteau, Dutrey, 
\& Simon}{1999}]{1999A&A...348..570G} Guilloteau S., Dutrey A., Simon M., 1999, A\&A, 348, 570 

\bibitem[\protect\citeauthoryear{Harris et al.}{2012}]{2012ApJ...751..115H} 
Harris R.~J., Andrews S.~M., Wilner D.~J., Kraus A.~L., 2012, ApJ, 751, 115 

\bibitem[Helou 
\& Walker(1988)]{1988iras....7.....H} Helou, G., \& Walker, D.~W.\ 1988, Infrared astronomical satellite (IRAS) catalogs and atlases.~Volume 7, p.1-265, 7,  

\bibitem[\protect\citeauthoryear{Hubickyj, Bodenheimer, 
\& Lissauer}{2005}]{2005Icar..179..415H} Hubickyj O., Bodenheimer P., Lissauer J.~J., 2005, Icar, 179, 415 

\bibitem[\protect\citeauthoryear{Hurley-Walker et 
al.}{2009}]{2009MNRAS.396..365H} Hurley-Walker N., et al., 2009, MNRAS, 
396, 365 

\bibitem[\protect\citeauthoryear{Kawabe et al.}{1993}]{1993ApJ...404L..63K} 
Kawabe R., Ishiguro M., Omodaka T., Kitamura Y., Miyama S.~M., 1993, ApJ, 
404, L63 

\bibitem[\protect\citeauthoryear{Koerner, Sargent, 
\& Beckwith}{1993}]{1993ApJ...408L..93K} Koerner D.~W., Sargent A.~I., Beckwith S.~V.~W., 1993, ApJ, 408, L93 

\bibitem[\protect\citeauthoryear{Krist et al.}{2005}]{2005AJ....130.2778K} 
Krist J.~E., et al., 2005, AJ, 130, 2778 

\bibitem[\protect\citeauthoryear{Leinert et 
al.}{1991}]{1991A&A...250..407L} Leinert C., Haas M., Mundt R., Richichi A., Zinnecker H., 1991, A\&A, 250, 407 

\bibitem[\protect\citeauthoryear{Lommen et 
al.}{2007}]{2007A&A...462..211L} Lommen D., et al., 2007, A\&A, 462, 211 

\bibitem[\protect\citeauthoryear{Mathis, Rumpl, 
\& Nordsieck}{1977}]{1977ApJ...217..425M} Mathis J.~S., Rumpl W., Nordsieck K.~H., 1977, ApJ, 217, 425 

\bibitem[\protect\citeauthoryear{McCabe, Duch{\^e}ne, 
\& Ghez}{2002}]{2002ApJ...575..974M} McCabe C., Duch{\^e}ne G., Ghez A.~M., 2002, ApJ, 575, 974 

\bibitem[\protect\citeauthoryear{Melis et al.}{2011}]{2011ApJ...739L...7M} 
Melis C., et al., 2011, ApJ, 739, L7 

\bibitem[\protect\citeauthoryear{Mugrauer 
\& Neuh{\"a}user}{2009}]{2009A&A...494..373M} Mugrauer M., Neuh{\"a}user R., 2009, A\&A, 494, 373 

\bibitem[\protect\citeauthoryear{Ohashi et al.}{1991}]{1991AJ....102.2054O} 
Ohashi N., Kawabe R., Ishiguro M., Hayashi M., 1991, AJ, 102, 2054 

\bibitem[\protect\citeauthoryear{Patnaik et 
al.}{1992}]{1992MNRAS.254..655P} Patnaik A.~R., Browne I.~W.~A., Wilkinson 
P.~N., Wrobel J.~M., 1992, MNRAS, 254, 655 

\bibitem[\protect\citeauthoryear{Perley 
\& Butler}{2013}]{2013ApJS..204...19P} Perley R.~A., Butler B.~J., 2013, ApJS, 204, 19 

\bibitem[\protect\citeauthoryear{Pi{\'e}tu et 
al.}{2011}]{2011A&A...528A..81P} Pi{\'e}tu V., Gueth F., Hily-Blant P., Schuster K.-F., Pety J., 2011, A\&A, 528, A81 

\bibitem[\protect\citeauthoryear{Pollack et 
al.}{1996}]{1996Icar..124...62P} Pollack J.~B., Hubickyj O., Bodenheimer 
P., Lissauer J.~J., Podolak M., Greenzweig Y., 1996, Icar, 124, 62 

\bibitem[\protect\citeauthoryear{Rice, Lodato, 
\& Armitage}{2005}]{2005MNRAS.364L..56R} Rice W.~K.~M., Lodato G., Armitage P.~J., 2005, MNRAS, 364, L56 

\bibitem[\protect\citeauthoryear{Rodmann et 
al.}{2006}]{2006A&A...446..211R} Rodmann J., Henning T., Chandler C.~J., Mundy L.~G., Wilner D.~J., 2006, A\&A, 446, 211 

\bibitem[\protect\citeauthoryear{Scaife et al.}{2008}]{2008MNRAS.385..809S} 
Scaife A.~M.~M., et al., 2008, MNRAS, 385, 809 

\bibitem[\protect\citeauthoryear{Scaife et 
al.}{2012}]{2012MNRAS.420.3334S} Scaife A.~M.~M., et al., 2012, MNRAS, 420, 
3334 

\bibitem[\protect\citeauthoryear{Simon 
\& Guilloteau}{1992}]{1992ApJ...397L..47S} Simon M., Guilloteau S., 1992, ApJ, 397, L47 

\bibitem[\protect\citeauthoryear{Song, Zuckerman, 
\& Bessell}{2003}]{2003ApJ...599..342S} Song I., Zuckerman B., Bessell M.~S., 2003, ApJ, 599, 342 

\bibitem[\protect\citeauthoryear{Tanaka, Himeno, 
\& Ida}{2005}]{2005ApJ...625..414T} Tanaka H., Himeno Y., Ida S., 2005, ApJ, 625, 414 

\bibitem[\protect\citeauthoryear{White et al.}{1999}]{1999ApJ...520..811W} 
White R.~J., Ghez A.~M., Reid I.~N., Schultz G., 1999, ApJ, 520, 811 

\bibitem[\protect\citeauthoryear{Wilner et al.}{2000}]{2000ApJ...534L.101W} 
Wilner D.~J., Ho P.~T.~P., Kastner J.~H., Rodr{\'{\i}}guez L.~F., 2000, 
ApJ, 534, L101 

\bibitem[\protect\citeauthoryear{Wilner et al.}{2005}]{2005ApJ...626L.109W} 
Wilner D.~J., D'Alessio P., Calvet N., Claussen M.~J., Hartmann L., 2005, 
ApJ, 626, L109 

\bibitem[\protect\citeauthoryear{Zwart et al.}{2008}]{2008MNRAS.391.1545Z} 
Zwart J.~T.~L., et al., 2008, MNRAS, 391, 1545 

\end{thebibliography}
\end{document}